# Plant-On-a-Disc (POD): A Phytofluidic platform enabling *In Situ* Root Analysis

Kaushal Agarwal[a], Sumit Kumar Mehta[b] and Pranab Kumar Mondal[a,c,1]

[a]School of Agro and Rural Technology,
Indian Institute of Technology Guwahati, Guwahati- 781039, India

[b]Department of Energy Engineering, SRM University-AP, Amaravati 522 240, Andhra Pradesh, India

[c]Microfluidics and Microscale Transport Processes Laboratory,
Department of Mechanical Engineering,
Indian Institute of Technology Guwahati, Guwahati- 781039, India

[1]Corresponding author; E-mail: pranabm@iitg.ac.in, mail2pranab@gmail.com (P. K. Mondal)
**ORCID:** 0009-0007-3152-5941 (Kaushal Agarwal), 0000-0001-8806-3767 (Sumit Kumar Mehta, sumitkumar.m@srmap.edu.in) and 0000-0002-9368-1532 (Pranab Kumar Mondal)

## Abstract

Phytofluidic platforms have enabled controlled studies of plant roots, however, most existing systems either impose geometric confinement without flow or introduce hydrodynamics in single-channel devices that limit throughput and disrupt downstream analysis. New experimental platforms are therefore needed to investigate how roots integrate mechanical confinement and hydrodynamic nutrient transport, two defining features of the rhizosphere that remain difficult to reproduce under controlled laboratory conditions. Here, we present the Plant-on-a-Disc (POD), a phytofluidic platform that enables the parallel cultivation of eight seedlings under controlled hydrodynamic conditions while allowing non-invasive, *in situ* multimodal analysis of the intact root-shoot system. The device is fabricated in PDMS using a cost-effective wire-drawing technique to generate radial microchannels that converge into a central sump beneath an optical window. This design enables sequential bright-field, fluorescence, and Raman measurements using a single microscope objective without disturbing neighbouring seedlings. Dimensionless transport analysis and finite-element modelling confirm that the radial architecture equalizes hydraulic resistance across channels, establishing creeping laminar flow with convection-dominated nutrient transport under physiologically safe shear conditions. Using *Brassica* seedlings, we show that hydrodynamic flow drives coordinated root responses across multiple scales. Roots grown in flow condition exhibit accelerated elongation, substantial ROS generation and anisotropic cortical cell expansion, accompanied by carotenoid signatures detected by Raman spectroscopy. Chlorophyll fluorescence measurements indicate stable photosystem II performance, while nitrogen uptake nearly doubles under flow conditions. Numerical simulations further reveal that hydrodynamic flow generates shear and surface stresses several orders of magnitude higher than those in stagnant conditions, providing a mechanistic basis for elongation-dominant growth.

## Introduction

Plant roots operate within a mechanically and chemically complex microenvironment,[1-3] where nutrient transport,[4-6] hydraulic flow,[7-9] and physical confinement[10,11] collectively regulate growth and metabolism.[12] In natural soils, roots rarely encounter static aqueous conditions.[1,13,14] Instead, they grow through tortuous pore networks, where micron-scale confinement and subsurface water flux generate spatial gradients of nutrients, oxygen, and mechanical stress.[15-20] These coupled physical cues influence root elongation, cellular organization, redox balance, and nutrient assimilation.[1,10,21-23] Understanding how roots integrate these transport and mechanical signals is therefore central to plant physiology and agricultural productivity.

Direct investigation of these processes in soil remains challenging, attributed primarily to the opaque nature and heterogeneous structure of the rhizosphere, which makes it difficult to perform experiments in a controlled environment.[24] Conventional approaches typically rely on soil extraction or destructive sampling, interrupting the root-shoot continuum and limiting real-time observation of dynamic responses.[25] To address these limitations, on-chip microfluidic systems have emerged as alternative soil-mimicking platforms, by creating transparent and precisely controlled environments for root growth and analysis.[26-29] The emerging field of phytofluidics integrates plant biology with microfluidic systems, to reproduce defined chemical and physical conditions while enabling continuous optical monitoring of roots within microchannels.[30-34] Such systems have enabled detailed studies of root tropisms, nutrient gradients, and root-microbe interactions with high spatial and temporal resolution. Despite these advances, most existing root-deployable on-chip systems remain limited in their ability to reproduce the coupled physical conditions, typically encountered in soils.[35-38] Many devices impose geometric confinement but operate under stagnant media, resulting in diffusion-dominated nutrient delivery.[22,28,39] Others introduce controlled flow but typically accommodate only a single seedling, limiting throughput and making comparative experiments labour-intensive.[31-33,35] In natural soils, roots simultaneously experience structural confinement within pore-scale architectures and dynamic hydraulic forces generated by subsurface nutrient flux. Collectively, these abiotic cues regulate root elongation, cellular organization, redox balance, and nutrient assimilation.[40-47] However, most current *in vitro* platforms isolate either confinement or flow, failing to capture their combined mechanistic influence on root growth and development. Besides, many existing devices require repeated decoupling from their microenvironment for capturing the morphographs of the growing roots.

This approach not only reduces experimental throughput but also introduces a significant risk of variability and cross-contamination during handling.

To address the aforementioned limitations, we propose a soil-mimicking on-chip fluidic device which we name as ‘Plant-On-a-Disc (POD)’ phytofluidic device. The proposed POD platform integrates soil-analogous confinement, controlled hydrodynamic flow, and *in-situ* multimodal analysis within a single device. The POD consists of eight radial PDMS microchannels converging into a common central sump. This radial architecture equalizes hydraulic resistance across channels, thereby enabling uniform flow conditions for multiple seedlings simultaneously. Besides, this multiplexed design eliminates the need for device decoupling, enhances experimental consistency, by maintaining identical environmental conditions across all channels, improves statistical reliability, and minimizes contamination risks. Consequently, the platform enables a more accurate estimation of *in-situ* root growth dynamics, an aspect that remains largely unexplored in the literature to date. We briefly discuss about the POD and its flow configuration in the forthcoming paragraph.

The POD platform is mounted on a rotatable stage that sequentially aligns each fluidic channel with the objective lens of an inverted microscope, enabling rapid imaging without disturbing the fluidic environment. A transparent PDMS-based POD system permits multimodal characterization, including bright-field microscopy, fluorescence imaging, and Raman spectroscopy, allowing the entire seedling to be analysed *in-situ* while preserving the root-shoot continuum. Although, dimensionless flow analysis underscores that the device operates within a creeping flow regime, we perform numerical simulations to estimate nutrient transport along the root surface employing convection-diffusion equations. Using this platform, we systematically dissect the contributions of confinement and nutrient flow to root morphogenesis by comparing stagnant microfluidic channels, which represent diffusion-limited pore environments with flow-induced microfluidic channels. Through an integrated numerical modelling-experimental approach that encompasses - the estimation of flow-induced stress on roots, morphometric analysis, Raman spectroscopy, reactive oxygen species (ROS) quantification, chlorophyll fluorescence measurements, and nitrogen uptake assays - we show that controlled flow within the POD induces coordinated morphological, biochemical, and physiological responses. Together, these results establish the POD as a soil-like phytofluidic platform that links hydrodynamics with plant physiology during the early stages of plant growth.

## Materials and Methods

### Device Fabrication

The POD, an integrated phytofluidic platform, was fabricated using polydimethylsiloxane (PDMS) for its excellent optical clarity, biocompatibility, and suitability for plant root visualization.[48,49] The disc has a diameter of 85 mm and a thickness of 7 mm, ensuring structural rigidity while maintaining compatibility with standard microscopy stages as shown in **Fig. 1a**. To fabricate the root growth phytofluidic channels embedded in the disc, a wire drawing method was employed.[50] These phytofluidic channels were designed to mimic pore pathways for growing roots. A circular petri plate having an inner diameter of 8.5 cm was used as the base mold. A hollow cylinder with an outer diameter 1.5 cm was placed at the centre of the petri plate which acted as the post-fabrication connection to the sump. Eight metal wires with a square cross-section of 1.0 mm × 1.0 mm and a length of 3 cm were radially arranged at equal intervals within the mold. The central hollow cylinder also served as a support structure to secure the wires in their designated positions. Pipette tips of measured length were fitted as plant inlet ports, posotioned 1cm from the disc periphery and oriented at an angled at of 45$^{o}$ relative to the wire axis. This arrangement enabled convenient seed placement and guided root growth along the flow direction. Each working channel extended 2.5 cm inward from the plant inlet port towards the central sump. Following the assembly of mold components, PDMS was poured into the setup ad cured at 45°C for 10 hours. After curing, the wires were withdrawn, leaving behind transparent phytofluidic channels with square cross-sections. Nutrient solution was supplied using a single syringe pump through flexible tubing and Luer connectors, each connected to the inlet ports of the phytofluidic channels. This arrangement allowed precise regulation of flow rates for each channel independently or synchronously. The outlets of all channels were connected to a central sump, which acted as a common waste reservoir to collect the outflowing nutrient solution.

### Plant Germination

*Brassica* seeds were surface sterilized following a standard procedure. The seeds were first washed with 70% ethanol for 30 s and subsequently immersed in 4% sodium hypochlorite solution any eliminate surface contaminants.[51,52] Following this, the seeds were rinsed 4-5 times with autoclaved distilled water to remove residual disinfectants. The surface-sterilized seeds were then placed on UV-sterilized filter paper arranged in a 5×5 pattern inside a sterile petri plate, which was subsequently sealed with Parafilm. The Petri dishes were incubated in darkness for 48 hours to facilitate germination.[53]

## Assembly of POD

The germinated seeds were carefully placed into the plant inlet ports of the POD, through the phytofluidic channels, as depicted in **Fig. 1b**. To establish controlled flow conditions, the POD was connected to a syringe pump using sterile syringes filled with full strength autoclaved MS medium, and baby feeding tubes under aseptic conditions. The seedlings were then allowed to grow within the channels of the POD inside a growth chamber maintained under standardized light (16:8 light/dark cycle), temperature (26°C), and relative humidity (75%) for 36 h. The syringe pump was operated at a flow rate of 0.4 ml/hr, previously identified as the optimal flow rate for *Brassica juncea*,[30,31] while control plants were grown under static (no-flow) conditions.

## Hydrodynamic Regime Characterisation and Dimensionless Analysis

The hydrodynamic regime within the Plant-On-a-Disc (POD) was characterised using dimensionless analysis under the experimental flow conditions employed during plant growth studies. Reynolds number (*Re*), Péclet number (*Pe*), and Sherwood number (*Sh*) were evaluated to assess the relative contributions of viscous forces, convective transport, and molecular diffusion within the phytofluidic channel of the POD. Two geometric flow configurations were considered to represent the range of possible flow conditions in the channel. In Scenario A, the channel was assumed to be fully open to flow without accounting for the root volume. The square channel cross-section (0.8 mm × 0.8 mm) corresponded to a flow area $A_A = 6.4 \times 10^{-7}\ \mathrm{m}^2$ and a hydraulic diameter $D_{h,A} = 8.0 \times 10^{-4}$ m. In Scenario B, the presence of a root of diameter $d_{root} = 0.4$ mm was considered, leaving a slot-like clearance with an effective average height of $h = 2.0 \times 10^{-4}$ m. The resulting effective average flow area was $A_B = 1.6 \times 10^{-7}\ \mathrm{m}^2$ with a corresponding hydraulic diameter $D_{h,B} = 4.0 \times 10^{-4}$ m. Fluid properties corresponding to an aqueous nutrient solution were used to calculate the values of the dimensionless parameters mentioned before in the analysis as follows: density $\rho = 1000\ \mathrm{kg\ m^{-3}}$, dynamic viscosity $\mu = 1.0 \times 10^{-3}$ Pa-s, and kinematic viscosity $\nu = 1.0 \times 10^{-6}\ \mathrm{m^2 s^{-1}}$. Nutrient diffusivity was assumed to be $D = 1.0 \times 10^{-9}\ \mathrm{m^2 s^{-1}}$.[54,55] The characteristic nutrient exchange length along the root was taken as $L = 2.5 \times 10^{-2}$ m. Experiments were conducted at a volumetric flow rate of $Q = 0.4\ \mathrm{mL\ h^{-1}}$, and the average fluid velocity was calculated using $U = Q/A$ for the respective flow areas.

We calculated both the Reynolds number, $Re = \frac{\rho U\ \ _h}{\mu}$,[56] and the Péclet number, $Pe = \frac{UD_h}{D}$, to assess the relative contributions of convection and diffusion.[57] Mass transfer at the root interface was characterised using the Sherwood number, estimated from the Leveque correlation for laminar flow in the entrance region, $Sh \approx \left(\frac{Re \cdot Sc \cdot D_h}{L}\right)^{1/3}$, where the Schmidt number is defined as $Sc = \nu/D$.[58] These dimensionless parameters were used to characterise the transport regime within the POD device under the applied flow conditions. For the flow configuration and fluid properties considered in this study, we obtain the values of several parameters as $Re \sim 1.39 \times 10^{-1}$, $Pe = 1.39 \times 10^{2}$, and $Sh \approx 3.03$.

**Modelling and Simulation of Root Morphology - Flow Interaction**

We perform simulations of the media flow - soft root structure interaction using COMSOL Multiphysics®, essentially to investigate the coupled hydrodynamic transport and mechanically induced stress distribution within a single phytofluidic channel containing a growing *Brassica* root. The computational geometry representing the root-channel assembly was reconstructed using SOLIDWORKS® based on high-resolution experimental micrographs of seedlings grown inside the POD microfluidic device. These micrographs served as references to reproduce the experimentally observed root morphology, including its diameter, spatial confinement, curvature, and orientation within the microchannel. The reconstructed three-dimensional CAD model consisting of the root and surrounding channel domain was exported as an STL file and imported into COMSOL Multiphysics® platform to define the computational domain. The computational domain consisted of two coupled regions: a fluid domain representing the aqueous nutrient medium flowing through the phytofluidic channel and a structural domain representing the root body subjected to hydrodynamic environment. The nutrient media solution was assumed to be a Newtonian, while the underlying flow was considered to be steady and incompressible flow assumptions. Hence, based on these assumptions, the governing equations for fluid flow were described by the continuity and Navier-Stokes equations:[59,60]

$$\nabla \cdot \boldsymbol{u} = 0 \quad [1]$$

$$\rho(\boldsymbol{u} \cdot \nabla)\boldsymbol{u} = -\nabla p + \mu \nabla^2 \boldsymbol{u} + \boldsymbol{f} \quad [2]$$

where $\boldsymbol{u}$ represents the velocity vector, $p$ the pressure field, $\rho$ the fluid density, and $\mu$ the dynamic viscosity of the nutrient solution. The letter, $\boldsymbol{f}$ demotes the gravitational body force experienced by the root represented as $-\rho g \hat{j}$. As already mentioned, the physical properties of the nutrient medium were approximated as those of water, with a density of $1000\ \mathrm{kg/m^3}$ and dynamic viscosity of $1 \times 10^{-3}\ \mathrm{Pa-s}$.

Boundary conditions were imposed to replicate the experimental operating conditions of the POD device. A uniform inlet velocity corresponding to the experimental volumetric flow rate (0.4 mL/hr) was imposed at the channel entrance. A zero-gauge pressure condition was imposed at the outlet ($p_{outlet} = p_{atm}$), while no-slip boundary conditions ($\boldsymbol{u} = 0$) were enforced along the channel walls and the root surface.

Mechanical stresses exerted by the flowing nutrient media solution were quantified using the following equation:[59-61]

$$\nabla(\boldsymbol{FS})^T = 0 \quad [3]$$

In Eq. 3, $\boldsymbol{F}$ and $\boldsymbol{S}$ denote the displacement gradient and the second Piola-Kirchhoff stress, respectively. The displacement field ($V$) can be represented in terms of displacement in gradient as follows:

$$\boldsymbol{F} = \boldsymbol{I} + \nabla V \quad [4]$$

where $\boldsymbol{I}$ denotes the identity tensor. The constitutive relationship between stress and strain for an isotropic elastic material and the second Piola-Kirchhoff stress is expressed in terms of Langrange-Green strain as given below:

$$\boldsymbol{S} = 2\mu_L \boldsymbol{\varepsilon} + \lambda_L t_r(\boldsymbol{\varepsilon})\boldsymbol{I} \quad [5]$$

where $\boldsymbol{\varepsilon}$ is the Langrange-Green strain and expressed as follows:

$$\boldsymbol{\varepsilon} = \frac{1}{2}(\nabla V + (\nabla \mathrm{V})^T) \quad [6]$$

The parameters, $\lambda_L$ and $\mu_L$ represent the Lamé elastic constants, which are related to the Young's modulus, $E$ and Poisson's ratio, $\nu$ as:[59]

$$\lambda_L = \frac{\nu E}{(1+\nu)(2\nu-1)}, \mu_L = \frac{E}{2(1+\nu)} \quad [7]$$

Further, the Cauchy stress tensor under the flow loading inside the root can be estimated by the following expression:

$$\boldsymbol{\sigma} = J^{-1}(\boldsymbol{FS}(\boldsymbol{F})^T) \quad [8]$$

The Jacobian for displacement gradient is represented by the symbol, $J$. To couple the deformation field with the flow field, the interfacial boundary condition is imposed following the continuity of resultant stress to surface in nutrient and root domains. The mathematical expression for this interfacial condition is represented as follows:

$$\boldsymbol{\sigma} \cdot \boldsymbol{n}_{r,\Omega} = (-p\boldsymbol{I} + \mu\nabla\boldsymbol{u}) \cdot \boldsymbol{n}_{N,\Omega} \quad [9]$$

Here, $\boldsymbol{n}$ represents the normal unit vector at the interface ($\Omega$) of root ($r$) and nutrient domain ($N$).

The internal mechanical state of the root is characterized using the Von Mises stress ($\sigma_v$), which is expressed as:[62]

$$\sigma_v = \sqrt{\frac{3}{2} S_{ij} S_{ij}} \quad [10]$$

Where, $S_{ij}$ denotes the deviatoric stress, representing the distortion-related component of the stress tensor ($\boldsymbol{S}$). The deviatoric stress tensor can be derived from the Cauchy stress tensor ($\boldsymbol{\sigma}$) as:[62]

$$\mathbf{S} = \boldsymbol{\sigma} - \sigma_m \mathbf{I} \quad [11]$$

Here, $\sigma_m$ is the mean (hydrostatic) stress, defined as:

$$\sigma_m = \frac{\sigma_{xx} + \sigma_{yy} + \sigma_{zz}}{3} \quad [12]$$

with $\sigma_{xx}$, $\sigma_{yy}$, and $\sigma_{zz}$ representing the normal stress components along the principal directions.[62]

To quantify the sectional averaged internal mechanical stress within the root volume, we performed the surface average of the von Mises stress as follows:

$$\sigma_{v,avg} = \frac{\iint \sigma_v dA}{\iint dA} \quad [13]$$

Here, $dA$ is the elemental cross-sectional area of root.

The elastic properties of the root tissue were defined using experimentally reported values for *Brassica juncea*, with a Young's modulus, $E = 4.1625$ MPa and Poisson's ratio $\nu = 0.49$.[31,32] The coupled governing equations were solved using the finite element method implemented in COMSOL Multiphysics®. The computational domain was discretized using a tetrahedral mesh, with additional mesh refinement near the root surface and channel walls to accurately resolve near-wall velocity gradients and stress variations under confined flow conditions. Boundary layer elements were incorporated at the nutrient-root interface to capture flow-induced shear effects. Linear shape functions were employed for the pressure and velocity fields in the fluid domain, while quadratic shape functions were used for the structural displacement field. The flow field was solved using the Generalized Minimal Residual (GMRES) iterative solver whereas the displacement field was solved using Parallel Direct Solver (PARDISO). Stationary simulations were performed using the aforementioned solver to compute the spatial distributions of velocity field, pressure field and displacement field. These simulations enabled quantitative characterization of the hydrodynamic loads and mechanical stresses experienced along the root surface under controlled phytofluidic flow and stagnant conditions. It is to be noted that for the stagnant condition (SiP) the load for the structural mechanics is derived from the hydrostatic pressure. We have done the grid independency test for the channel with nutrient flow and stagnant conditions, resulting in optimized mesh sizes of 5,04,076 and 4,28,496 elements, respectively.

**Microscopy and Anatomy**

The root morphology of seedlings grown in the POD under flow and stagnant conditions were observed using brightfield microscopy (Bestscope™) at regular time interval of 2 hr over a total experimental duration of 36 hrs. It is to be noted that POD is compatible with the stages of both upright and inverted microscopes. The roots were also examined using fluorescence microscopy to analyse changes in longitudinal cellular morphology under both flow and stagnant conditions. Fluorescent staining through propidium iodide (roots dipped in PI for 30 s) enabled clear visualization of cell walls thereby facilitating detailed morphological analysis. Also, free-hand transverse sections of the roots were prepared stained with safranine, and observed under brightfield microscope for anatomical characterization. Quantitative measurement of root length, and average cell dimensions were performed using the open-source software ImageJ.

**Raman Spectroscopy**

Following the completion of the 36-hr experimental period, Raman spectroscopy (Make: Horiba Jobin Vyon, Model LabRam HR) was performed on roots grown under both flow and stagnant conditions using a 488 nm diode laser (acquisition time 10s). The obtained Raman spectra were baseline corrected and analysed using OriginPro® software to identify peaks associated with specific metabolites. Flow induced stress experienced by the roots, also referred to as abiotic stress, leads to the accumulation of different metabolites and alters biochemical regulation in the growing roots. Differences in peak intensity and distribution were interpreted as indicators of flow-induced biochemical modifications.

**Reactive Oxygen Species (ROS): Fluorescent Microscopy and Quantification**

After 36 h of growth, intracellular reactive oxygen species (ROS) in the roots were visualized using 2′,7′-dichlorodihydrofluorescein diacetate (DCFH-DA). The stained roots were imaged using a fluorescence microscope (Make: Nikon, Model: H600L) under identical imaging conditions for all treatments. The fluorescence intensity was quantified using ImageJ software. A line was drawn along the root starting from the tip to the elongation region, and the fluorescence intensity was measured after background subtraction. The intensity profiles were used to compare ROS levels in roots grown under different experimental conditions, *viz*., Flow in POD (FiP), Stagnant in POD (SiP) and Stagnant in petri plate (SiPP), as elaborated later on in the upcoming section.

**Nitrogen Uptake Assay**

A nitrate-based nutrient solution (MS media) was supplied to the roots via syringe pump. After a fixed growth durations of 36 hours, root tissues were analysed to determine total nitrogen content by the Kjeldahl method using a Total Kjeldahl Nitrogen (TKN) analyser (Kelpus-Distyl EM VA, Pelican Equipment, India).[31,32,63]

## Results and Discussion

To mechanistically dissect the influence of confinement and flow environment, three experimental conditions were established: (i) Flow in POD (FiP): roots grown within the phytofluidic channel under controlled nutrient flow, representing structural confinement combined with hydrodynamic stimulation. (ii) Stagnant in POD (SiP): roots grown within the phytofluidic channel without flow, representing confinement under diffusion-limited nutrient conditions. (iii) Stagnant in petri plate (SiPP): roots grown without confinement and without flow, serving as the baseline condition. Comparison of SiP versus SiPP isolates the effect of architectural confinement, whereas comparison of FiP versus SiP isolates the effect of hydrodynamic shear stress and advective nutrient renewal. This structured experimental design enables clear differentiation between the effects of geometrical confinement and flow-mediated influences on root morphology and metabolism.

### Device Architecture and Experimental Configuration: A High-Throughput Phytofluidic Platform

The Plant-on-a-Disc (POD) phytofluidic platform was fabricated using polydimethylsiloxane (PDMS) due to its optical transparency, biocompatibility, and suitability for root imaging, as shown schematically in **Fig. 1a**.[48,49] The disc has a diameter of 8.5 cm and thickness of 7 mm, ensuring mechanical stability and compatibility with standard microscopy stages for real-time root growth observation. The POD consists of eight symmetrically arranged radial microchannels embedded within the PDMS disc. Each microchannel contains an independent set of ports: a nutrient inlet port connected to the syringe pump through a baby feeding tube, a plant inlet port for seedling placement, and an outlet port connected to a common sump. The channels were fabricated using a wire-drawing technique,[50] as detailed in the *Materials and Methods section*. This method produced well-defined square microchannels with consistent geometry and leak-free sealing, optimized for laminar flow and guided root development. After establishing the channel geometry, the POD was integrated with essential components for investigating root morphometric, mechanical, and hydrodynamic parameters. The overall working setup of POD device is shown in **Fig. 1 b (i-iv)**.

Each nutrient inlet port was connected to a 10 mL sterile syringe (NIPRO, India) preloaded with Murashige and Skoog (MS) nutrient medium (PT021, HIMEDIA®, India), a standard formulation for plant tissue culture, as depicted in **Fig. 1b**. The nutrient solution, which behaves as a Newtonian fluid with rheological properties similar to water, was delivered into the device through sterile flexible tubing (500 mm in length and 2.70 mm in diameter). Flow was precisely regulated using a programmable multi-channel syringe pump (New Era Pump Systems, Inc., USA), allowing independent and synchronized flow at 0.4 mL/hr to a central sump (**Fig. 1b**) that functioned as a common reservoir for effluent collection. To maintain aseptic conditions throughout the experiment and to support optimal seedling growth, the entire setup —including the POD device, nutrient supply system, and sump - was housed within a UV-sterilized, closed plant growth chamber under controlled environment conditions of 26 °C temperature, 75% humidity, and a 16 hr light/8 hr dark photoperiod.

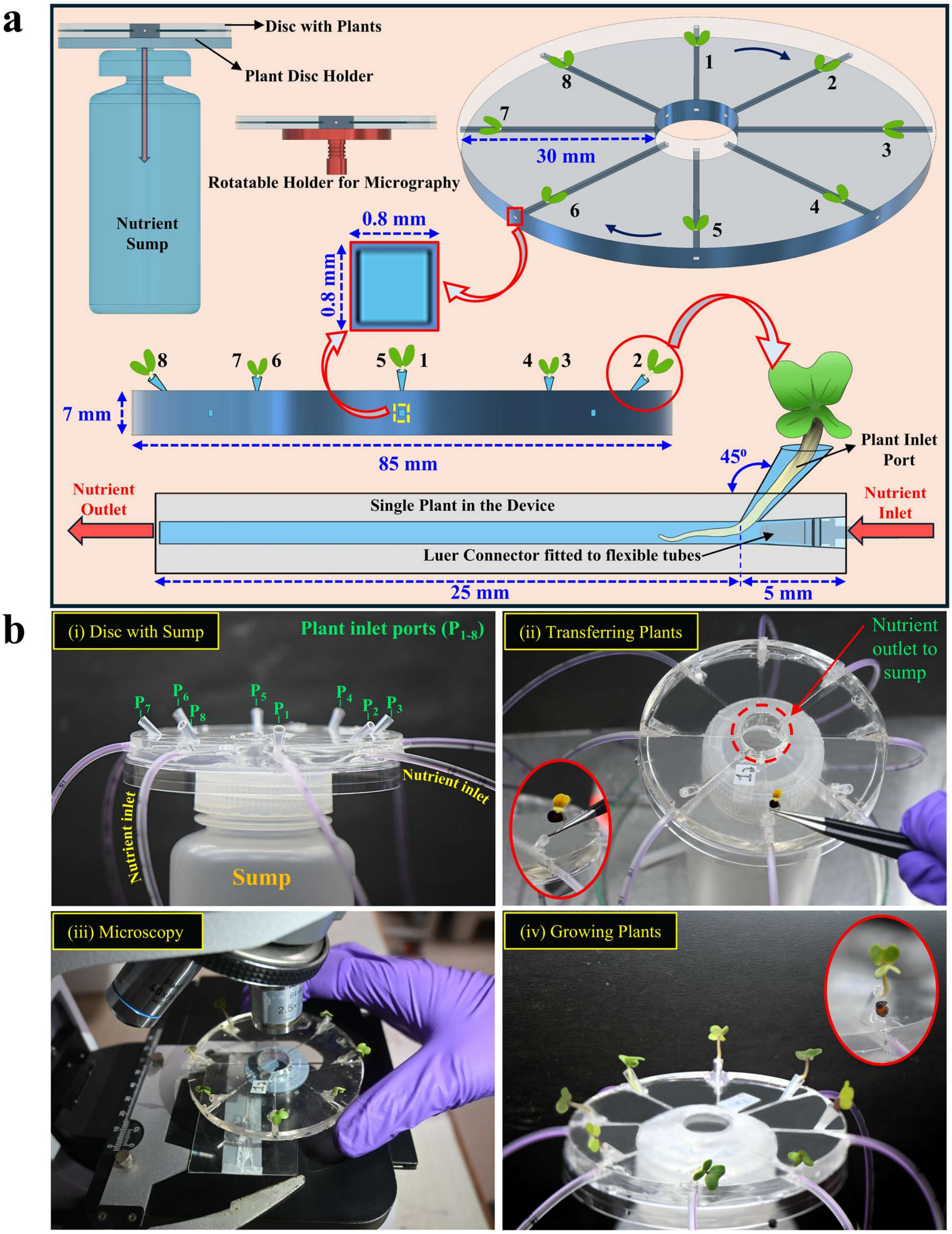

**Figure 1**: Design, fluidics, and handling workflow of the Plant-on-a-Disc (POD) plant flow platform. (a) Schematic representation of the POD showing a cast-PDMS disc hosting eight radial micro-channels that converge into a central sump. The disc is clamped to a rotatable holder mounted on an inverted microscope and sits atop a 50 mL nutrient reservoir ("sump") that serves simultaneously as outlet and degassing trap. Each channel has two ports: a rear nutrient inlet (Luer connector) and a forward plant inlet where a pre-germinated seedling is inserted; the channel terminates in the sump, establishing a through-flow path. (b) Handling workflow of POD. Photograph of disc on the sump. showing micro-tubing lines (nutrient inlets) attached to the disc. A 2-day-old *Brassica* seedling was gently placed root-first into the inclined plant port; capillary fit prevents leakage while allowing the shoot to emerge. The assembled disc is rotated beneath a 2.5 × long-working-distance objective for root microscopy without disturbing the remaining channels. After 36 hrs, all roots extend into their individual micro-channels while the cotyledons remain above the disc surface, enabling simultaneous flow and in-situ micrography and measurements.

The POD phytofluidic platform offers a high-throughput and uniform-environment solution for controlled root growth experiments. In contrast to earlier standalone single-channel devices, which accommodated only one seedling at a time, required prolonged operational periods, and often suffered from inconsistent microenvironments, the POD can simultaneously host eight seedlings under identical and precisely regulated conditions. This arrangement ensures reproducibility, reduces analysis time, and minimizes manual intervention. The compact, disc-shaped configuration is mechanically robust and fully compatible with standard microscopy stages, allowing the device to be smoothly rotated under the objective lens for multi-angle, high-resolution imaging of all seedlings in quick succession. This rotatable design not only accelerates imaging throughput but also eliminates the need for repeated sample repositioning, thereby maintaining stable growth conditions during the entire period of experimental observation. The handling and positioning of the POD device during microscopic imaging are demonstrated in Movie S1.

Jointly, these features make the POD efficient, easy to operate, and ideally suited for comparative root phenotyping and flow-induced plant studies. Building on these capabilities, the experimental workflow began with the careful transfer of germinated *Brassica juncea* seedlings into the custom-engineered Plant-on-a-Disc (POD) phytofluidic platform for controlled root growth studies (**Fig. 1a** and **1b**).

**Design Rationale: Mimicking Soil Hydraulics, Ensuring Soil-Analogous Design Framework and Hydrodynamic Validation**

The design philosophy of the Plant-on-a-Disc (POD) was rooted in a fundamental question: ***how can an in vitro platform reproduce the essential hydraulic and mechanical cues naturally experienced by roots in soil, while simultaneously preserving the optical accessibility needed for high-resolution, multimodal analysis?*** To address this challenge, the Plant-on-a-Disc (POD) was designed to replicate the hydraulic and mechanical microenvironment of soil while maintaining analytical transparency and experimental controllability. In natural soil, roots grow within a porous and mechanically confined environment characterized by heterogenous fluid flow, spatial gradients of nutrients, and dynamically varying stress distributions. These coupled factors collectively regulate root elongation, cellular organization, mechanosensory signalling, and nutria
nt uptake. Conversely, most existing root-on-chip systems operate either under static conditions or within simplified single-channel flow configurations, thereby failing to capture the complex interplay among confinement, hydrodynamics, and stress-mediated root responses.

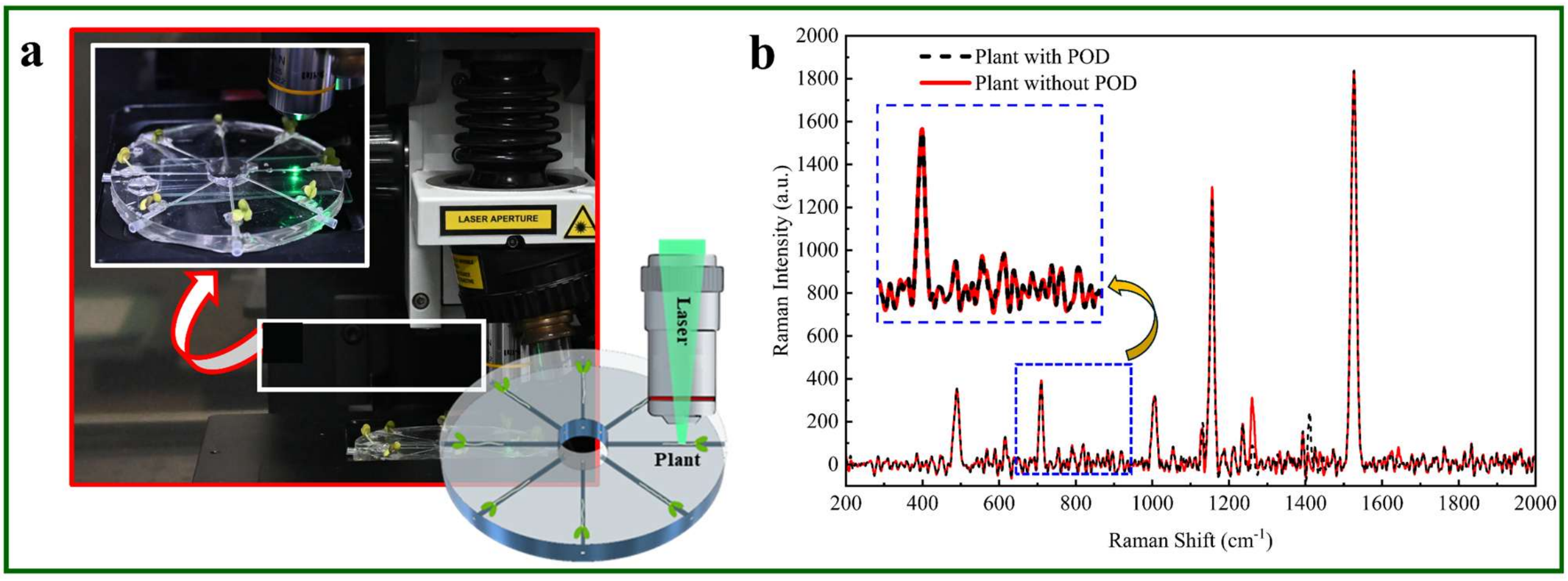

**Figure 2**: Raman mediated neutrality of the POD (a) Photograph of the Plant-on-a-Disc (POD) mounted on the confocal Raman microscope; the inset highlights the optically transparent PDMS window enabling non-invasive spectral acquisition without disturbing the root-channel interface. (b) Whole-seedling Raman spectra comparing plants grown with and without the POD confirm spectral neutrality of the PDMS housing. The inset provides a magnified view of the selected spectral region, highlighting the overlap between the two spectra.

To address these limitations, the POD incorporates eight radial microchannels converging into a central sump, thereby ensuring uniform pressure distribution and similar hydraulic conditions across all channels. (**Fig. 1**). This architecture ensures that every root experiences a comparable hydraulic regime, thereby eliminating effectively minimizing inter-sample variability. In addition, the radial configuration allows sustained flow while maintaining physiologically safe shear conditions suitable for long-term growth and observation.

To determine whether this conceptual design realistically mimics soil-like hydraulic scenarios, we employed a two-tiered validation strategy, from both experimental and numerical perspectives. The first validation strategy was experimental and focused on an important concern: ***whether the device itself could introduce artefacts into optical or biochemical measurements.*** Since the POD was specifically designed to facilitate *in situ* analytical techniques such as Raman spectroscopy, where spectral integrity is crucial, it was essential to ensure that the PDMS housing, microchannel geometry, and optical window alignment did not distort the molecular fingerprints of the roots. To evaluate this, Raman spectra of *Brassica* roots were collected directly through the PDMS optical window (**Fig. 2a**) and subsequently compared with spectra obtained after gently removing the roots from the device (**Fig. 2b**). The spectra obtained through the PDMS window closely matched those recorded outside the device, with no observable spectral distortion or additional peaks (**Fig. 2b**). This finding was significant because it confirmed that the POD itself does not interfere with biochemical measurements. Consequently, any spectral changes observed in subsequent experiments can be attributed to genuine flow- or confinement-induced mechanobiological responses of the plant roots rather than artefacts arising from the devise architecture.

After establishing biocompatibility through Raman spectral results, a dimensionless analysis was performed as a numerical cross-check of the hydrodynamic regime and mass transfer characteristics within the phytofluidic channel. The Reynolds number ($Re \sim 1.39 \times 10^{-1}$) confirmed a creeping-flow regime within the channel of the POD, characteristic of viscous-dominated transport similar to that observed in soil pores. The Péclet number ($Pe = 1.39 \times 10^{2}$) indicated a strongly convection-dominated transport regime while, the Sherwood number ($Sh \approx 3.03$) suggested an approximately threefold enhancement in mass transfer compared to pure diffusion. Collectively, these scaling results suggest that stagnant channels replicate diffusion-limited microporous regions of soil, while flow-induced channels simulate the pore-flushing associated with subsurface seepage. Thus, even in the absence of flow, the POD provides a realistic confinement baseline, whereas under flow condition, it reproduces soil-like advective renewal. These two complementary regimes may subsequently explain the observed differences in root remodelling, biochemical responses, and nutrient uptake. The evaluated dimensionless parameters (Re, Pe, and Sh), and their magnitudes not only support the numerical predictions but also position the POD within a soil-relevant and physiologically favourable hydrodynamic window. Collectively, the system operates in a regime characteristic of natural soil pore environments, where creeping flow and intermittent advective renewal coexist with geometric confinement.

Concurrently, the numerical scaling analysis and the experimental validations demonstrate that the POD successfully mimics soil-like hydraulics while ensuring both hydrodynamic and analytical integrity. From the hydrodynamic perspectives, the platform generates a microenvironment characterized by gentle but effective shear, creeping laminar flow, and convection-dominated transport, conditions that are known to stimulate elongation and nutrient uptake in roots. From an analytical perspective, it preserves the fidelity of Raman optical signals (**Fig. 2**), providing a clear and artefact-free window window into root biochemistry. By uniting these two aspects i.e, soil-inspired hydraulic mimicry and artefact-free analytical capability, the POD establishes itself as both a physiologically safe and methodologically rigorous platform. This combination forms the foundation for the reliability of all subsequent observations related to root morphology, cellular organisation, metabolite enrichment, photosynthetic performance, and nitrogen assimilation, described in the later sections.

**Flow- and Confinement-Driven Root Morphology and Cellular Remodelling**

With the hydrodynamic and analytical validation of the POD established, the central question was ***whether this engineered microenvironment could actually drive root remodelling in a manner analogous to soil-imposed mechanical and nutritional stresses.*** In natural soil, elongating roots are not free-floating structures; instead, they grow through tortuous pore spaces, where frictional resistance, localized confinement, and variable water availability combine to regulate extension and tissue organization. To emulate these conditions, root growth investigated under two geometrically confined POD channels: (i) stagnant nutrient-filled channels, where confinement acted in the absence of bulk fluid motion, and (ii) channels where roots were simultaneously exposed to hydrodynamic shear and continuous nutrient replenishment. This bipartite comparison was essential to separate the contributions of confinement alone from those arising due to hydrodynamic stress or flow-induced mechanical stress.

The root length data show a clear difference between the confined growth under FiP and SiP conditions, with roots in the FiP condition growing faster than those in the SiP condition throughout the 36 hrs period (**Fig. 3a**). This trend is also visible in the sequential root images, where roots under FiP show greater elongation than those under SiP (**Fig. 3b**). The insets in **Fig. 3b** show the microscopy setup used for imaging, the top view of the Plant-on-a-Disc (POD) platform, and a magnified view of the growing root inside the confined phytofluidic channel. These images illustrate the experimental configuration used to monitor root growth throughout the study. Roots exposed to continuous nutrient flow exhibited consistently higher elongation rates throughout the experiment period. After 36 h, nutrient flow-mediated roots reached lengths of approximately 24.63 ± 0.57 mm, whereas roots maintained in stagnant channels grew only to 15.55 ± 0.48 mm. A representative time-lapse sequence illustrating root growth within the POD platform is provided in Video S2. Notably, the root growth behaviour observed under stagnant nutrient conditions within the phytofluidic channel closely resembled that typically observed -in conventional petri plate culture, where nutrient replenishment and mechanical simulation are minimal. Contrarily, the introduction of flow within the confined phytofluidic channel, comparable to that of the water movement (flow) through soil pore networks,[64,65] markedly enhanced axial root extension. These observations highlight that root elongation is governed not only by geometric confinement but also by the combined effects of continuous nutrient renewal and flow-induced mechanical factors. Together, the results suggest that confinement and hydrodynamic conditions interact to modulate the baseline elongation behaviour of roots within the POD microenvironment.

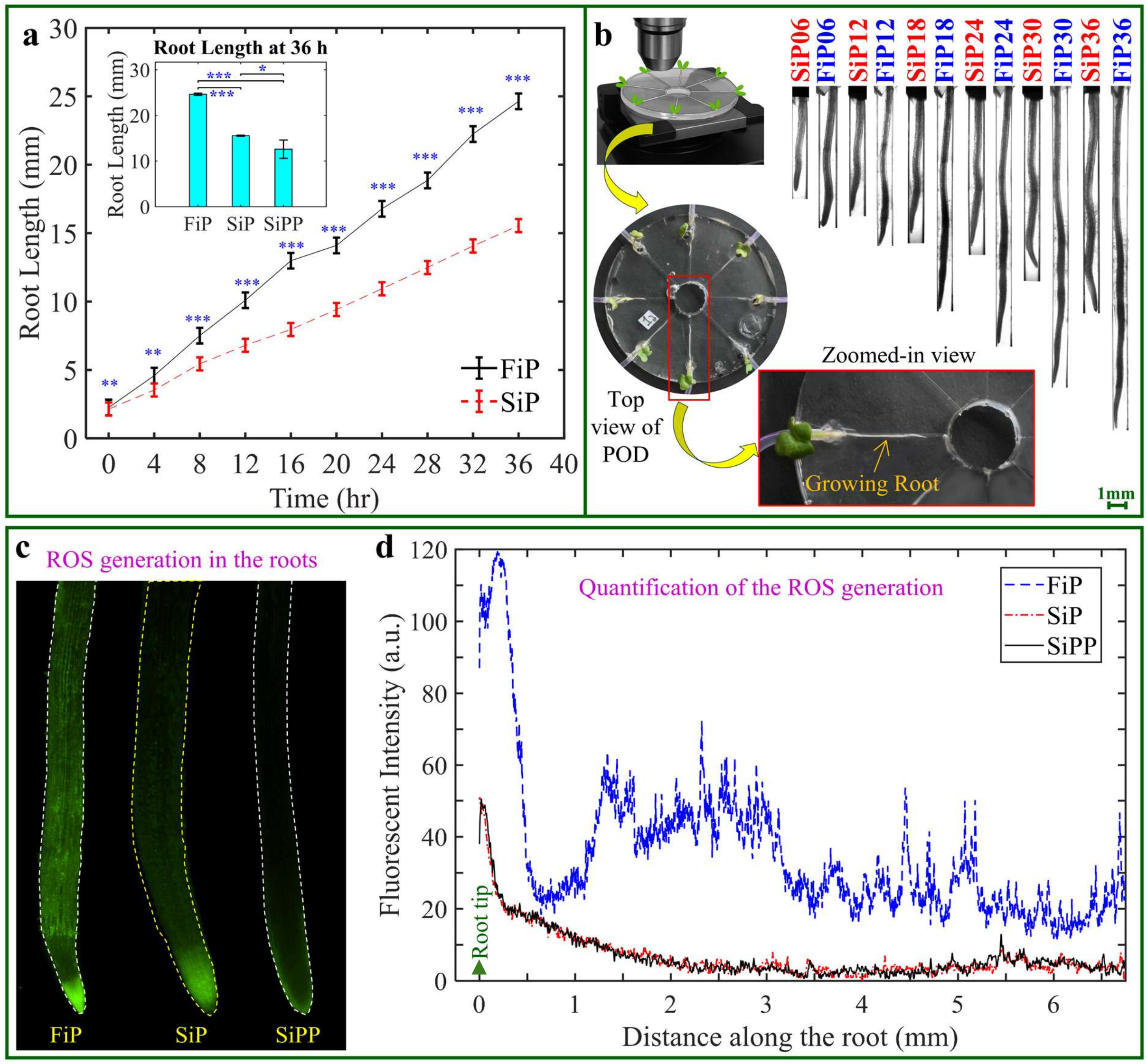

**Figure 3**: Hydrodynamic flow enhances root elongation in the phytofluidic channel. (a) Time-dependent root length of seedlings under hydrodynamic flow and stagnant conditions over 36 h. Roots exposed to continuous flow exhibit significantly faster elongation compared to stagnant conditions, as reflected by the steeper growth trajectory. Inset: Root length at 36 h for FiP, SiP, and SiPP conditions, showing that roots grown under flow reached the highest length. (b) Representative sequential root images at different time points (6 hours interval, upto 36 hours) under SiP and FiP conditions, visually illustrating the greater elongation under flow conditions. Insets: Microscopy setup, top view of the Plant-on-a-Disc (POD) device, and a magnified view of the growing root inside the phytofluidic channel. (c) Representative fluorescence images showing reactive oxygen species (ROS) accumulation in roots under FiP, SiP, and SiPP conditions. (d) Fluorescence intensity profiles along the root length, showing substantial ROS accumulation under FiP condition than that of SiP and SiPP condition. Data represent mean ± SD (n = 3). Statistical significance between treatments was determined using two-way ANOVA followed by Tukey's post-hoc test (**$p < 0.01$, ***$p < 0.001$).

To examine whether the different growth conditions influenced the physiological state of the roots, reactive oxygen species (ROS) generation was analysed using fluorescence imaging. Representative fluorescence images show much stronger ROS signals in roots grown under the FiP condition than in the SiP and SiPP conditions (**Fig. 3c**). The fluorescence intensity profile measured along the root length also exhibits a much higher signal under FiP, whereas the SiP and SiPP roots maintained low and nearly overlapping intensity values throughout the measured region (**Fig. 3d**). It may be mentioned here that though a subtle yet noticeable difference in ROS accumulation is witnessed in **Fig. 3c**, depicting the fluorescence profiles for FiP, having a much higher ROS generation, leads to the overlapping of the profiles for SiP and SiPP conditions (cf. **Fig. 3d**). The higher ROS level under hydrodynamic flow indicates an active physiological response to the flow environment. Despite this increase in ROS, roots under FiP exhibited the highest elongation (**Fig. 3a**). This observation underscore that a relatively higher ROS generation in FiP is accompanied with normal growth and cellular remodelling rather than growth inhibition. These results indicate that hydrodynamic flow not only enhances root elongation but also modifies the physiological activity of the growing root. These qualitative evidences of eustress is also substantiated by quantitative perspectives, including Raman spectra analysis and numerical simulations, as discussed in the forthcoming sections.

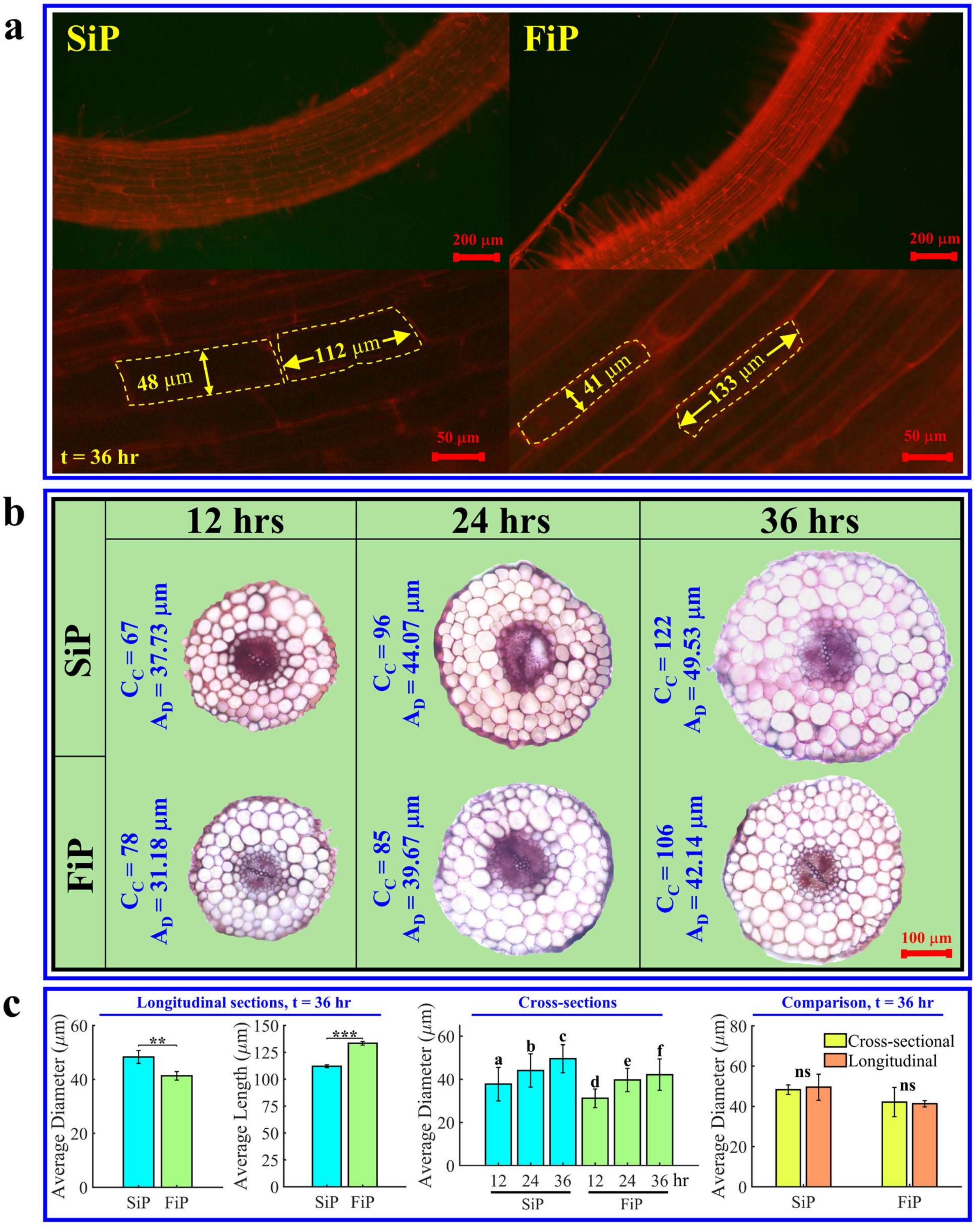

**Figure 4**: Hydrodynamic regulation of root elongation and cortical remodelling in the POD platform. (a) Propidium iodide-stained longitudinal fluorescence images (4× and 10×) reveal pronounced anisotropic cell expansion under flow. Cortical cells elongate longitudinally (133 ± 1.69 µm in roots under flow condition, FiP vs 112 ± 1.32 µm in roots under stagnant condition, SiP) while transverse diameter decreases (41 ± 1.56 µm vs 48 ± 2.38 µm), indicating shear-biased strain redistribution and elongation-dominant growth. Statistical significance was determined using one-way ANOVA followed by Tukey's post-hoc test (** $p < 0.01$ for diameter, *** $p < 0.001$ for length). (b) Safranine-stained transverse sections obtained 5 mm behind the root tip at 12, 24, and 36 h show time-resolved cortical

reorganization. Under stagnant conditions, cortical cell number increases progressively ($C_c$ = 67, 96, 122), consistent with radial thickening. In contrast, flow-exposed roots exhibit moderated radial proliferation ($C_c$ = 78, 85, 106) and a more compact cortical lattice, with consistently smaller root diameters across time ($A_D \approx 31.18$-$42.14$ µm under flow vs 37.73-49.53 µm under stagnant conditions). Statistical significance was determined using one-way ANOVA followed by Tukey's post-hoc test revealing a statistically significant difference among the experimental groups ($p < 0.001$). (c) Quantitative measurements of cortical cell dimensions and root diameter. At 36 h, FiP roots exhibit reduced cell diameter and increased cell length compared to SiP roots, indicating elongation-dominant growth. Root diameter increases with time under both conditions but remains lower under FiP. No significant difference was observed between diameters measured from longitudinal and transverse sections at 36 h (ns). Data are mean ± SD; different letters indicate significant differences ($p < 0.05$).

The temporal micrographs of root growth kinetics further emphasize the morphological divergence between the two experimental conditions over the course of development (**Fig. 3b**). At every observed time point between 6 and 36 h, roots exposed to continuous flow exhibited visibly greater axial elongation compared to those growing under stagnant conditions. These sequential observations confirm that the divergence in root length is not merely a late-stage effect but rather emerges progressively during the early development as hydrodynamic effects continuously modify the root microenvironment.

At the cellular level, fluorescence microscopy reveals distinct patterns of cortical cell expansion (**Fig. 4a**). Under flow conditions, cells exhibit pronounced axial elongation, are longer (133 ± 1.69 µm) and narrower (41 ± 1.56 µm), indicating anisotropic elongation in which cellular growth is preferentially directed along the longitudinal axis of the root (**Fig. 4a**). In comparison, under stagnant conditions, cells are shorter (112 ± 1.32 µm) and wider (48 ± 2.38 µm), resulting in thicker and less elongated tissue architecture (**Fig. 4a**). One-way ANOVA followed by Tukey's post-hoc test (**Fig. 4c**) confirmed significant differences in both diameter (**$p < 0.01$ for diameter) and length (***$p < 0.001$). This is consistent with cross-sectional measurements at the end of 36 hrs (**Fig. 4b**), where roots grown under flow condition consistently exhibit smaller overall diameters ($A_D \approx 42.14$ µm) compared to roots grown under stagnant condition ($A_D \approx 49.53$ µm). One-way ANOVA revealed a statistically significant difference among the experimental groups ($p < 0.001$). Tukey's HSD post-hoc analysis confirmed significant pairwise differences between multiple groups, indicating that both treatment condition and duration significantly influenced cell diameter (**Fig. 4c**).

Furthermore, cortical cell diameters measured from cross-sectional and longitudinal sections showed no significant difference for either stagnant or flow conditions (**Fig. 4c**, $p > 0.05$), confirming the consistency of the measurements across both sectioning orientations. These agreements confirm that reduced transverse cell width under flow translates to suppressed radial expansion at the tissue level, whereas increased cell width under stagnant conditions manifests as macroscopic root thickening. The shear stress generated by flowing nutrient media likely promotes strain redistribution along the longitudinal axis, facilitating elongation while suppressing excessive radial swelling. In channels with stagnant nutrient media, however, the absence of tangential mechanical stimulation allows growth strain to redistribute more isotropically, resulting in broader cortical cells and reduced axial extension. Cross-sectional histological analysis further reveals how these hydrodynamic environments influence internal tissue organization over time (**Fig. 4b**). At the early developmental stage of 12 h, roots grown in flow-mediated condition show a slightly higher average cortical cell count ($C_c = 78$) compared with stagnant roots ($C_c = 67$), suggesting that continuous nutrient supply and mild mechanical stimulation initially promote active cortical cell proliferation. However, as development progresses, this trend reverses. By 24 h and 36 h, stagnant roots display greater cortical cell numbers ($C_c = 96$ and 122) relative to roots grown in flow-mediated condition ($C_c = 85$ and 106). Statistical significance was determined using one-way ANOVA followed by Tukey's post-hoc test revealing a statistically significant difference among the experimental groups ($p < 0.001$).

Thus, a clear structure-function relationship emerges: flow conditions favour elongation-dominated growth (fewer, longer, and narrower cells: thinner but longer roots), while stagnant conditions favour proliferation- and swelling-dominated growth (more, shorter, and wider cells: thicker but shorter roots). This difference arises from transport conditions: diffusion-limited environments create depletion of nutrients and promote radial growth for roots under stagnant conditions, while flow conditions maintain uniform nutrient availability and supports elongation for the roots. Such a configuration is mechanically efficient for growth within a flow microenvironment, which is typical to soil,[64,65] and aligns with the observed enhancement in overall root elongation.

Overall, the observations from macroscopic growth measurements and microscopic structural analyses show that nutrient transport within the POD actively governs both the direction and structural strategy of root development (**Figs. 3 and 4**). From the foregoing discussion, it can be inferred here that, stagnant conditions favour radial thickening, while flow promotes axial elongation through continuous nutrient supply and shear effects. These architectural differences emerge across multiple biological scales, from whole-root elongation to cellular geometry and internal tissue organization.

Importantly, the mechanically induced morphological changes observed in the POD correspond closely with the physiological and biochemical responses discussed in subsequent sections. Enhanced elongation under flow is associated with increased nitrogen uptake and structural reinforcement of root tissues. Thus, in the POD system, fluid flow acts not only as a transport mechanism but also as a critical physical regulator of root morphogenesis and adaptive growth.

**Biochemical Reprogramming Under Confinement and Flow: Raman Analysis**

While morphological remodelling provides a visible indication of hydrodynamic stimulation, a more sensitive measure of root adaptation lies in biochemical signatures, which reflect changes in cell wall composition, pigments, and overall metabolic state. To capture these biochemical responses, we used *in situ* Raman microspectroscopy as a label-free and non-destructive technique to monitor key biochemical components such as carotenoids, cellulose, proteins, and metabolic intermediates. Prior to Raman analysis, spectral neutrality of the POD was confirmed by comparing spectra obtained through the PDMS window with those from roots outside the device, showing near-identical overlap (**Fig. 2**). This allowed reliable acquisition of both root and shoot spectra under three conditions discussed earlier, *viz*., SiPP, SiP, and FiP (**Fig. 5**).

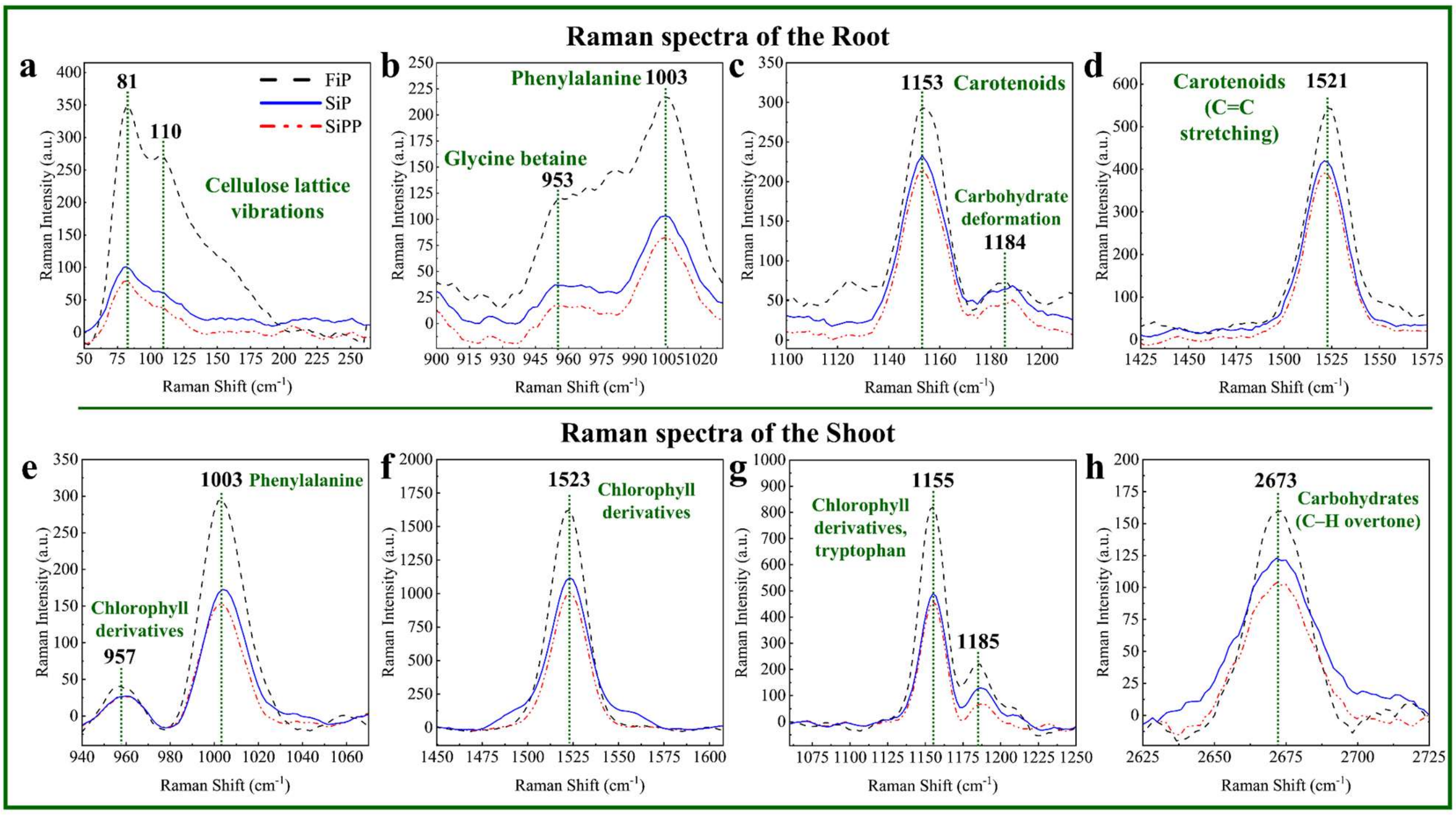

**Figure 5**: *In situ* Raman microspectroscopy reveals graded biochemical enhancement under confinement and flow in the POD platform. (a-d) Root spectra under flow in POD (FiP), stagnant in POD (SiP), and stagnant in petri plate (SiPP) conditions display a clear intensity hierarchy (FiP > SiP > SiPP) across phenylalanine, carbohydrate/cellulose, and carotenoid bands, indicating progressive wall reinforcement and antioxidant enrichment. (e-h) Shoot spectra show similar enhancement under flow, with stronger chlorophyll-, phenylalanine-, tryptophan-, and carbohydrate-associated bands compared with stagnant and petri plate controls, indicating systemic metabolic responses associated with root-level hydrodynamics.

These conditions separate the effects of confinement and flow. Seedlings grown under SiPP experience a minimally constrained, hydrostatic/normal stress in a diffusion-dominated aqueous environment. In SiP, roots grow inside the confined POD channels, so they experience confinement together with hydrostatic/normal stress, but without any flow or advective nutrient transport, allowing isolation of biochemical responses associated with pore-scale confinement. In FiP, roots were grown in the same confined channels under controlled flow environment, adding shear stress and convective nutrient transport to the existing confinement and hydrostatic pressure. Thus, the stepwise progression (SiPP → SiP → FiP) comparison allows clear identification of biochemical responses to confinement and flow in the Raman spectra (**Fig. 5**).

Raman spectra of roots (**Fig. 5a-d**) show a graded biochemical increase across conditions (FiP > SiP > SiPP). Low-frequency peaks at ~81 and ~110 $cm^{-1}$ correspond to lattice vibrations associated with deformation of the cellulose microfibril network (**Fig. 5a**). Under SiP, peaks at ~953 $cm^{-1}$ (glycine betaine)[66] and ~1003 $cm^{-1}$ (phenylalanine)[67] increase compared to SiPP (**Fig. 5b**). The ~953 $cm^{-1}$ glycine betaine[66] band indicates accumulation of an osmoprotectant involved in stress regulation, including hormonal balance, reactive oxygen species detoxification, and ion homeostasis.[68] Peaks at ~1153 $cm^{-1}$ (C-C stretching of carotenoids) and ~1184 $cm^{-1}$ (carbohydrate deformation) also increase under confinement (**Fig. 5c**).[69] Increased carotenoid bands indicate enhanced antioxidant pigments that buffer ROS generated during mechanosensory activation and stabilize membranes, while the carbohydrate deformation band suggests enhanced structural polysaccharide organization within the root cell walls. The carotenoid-associated peak at ~1521-1523 $cm^{-1}$ further increases under confinement (**Fig. 5d).** [69] Together, these trends suggest increased osmoprotection, antioxidant activity, and structural organisation under confinement.

Having a closer look at **Fig. 5a-d**, it is witnessed that for FiP, these signals intensify further, confirming enhanced biochemical activity under flow. Increased carotenoid, phenylalanine, and carbohydrate bands indicate improved antioxidant capacity, metabolic activity, and cell-wall organisation under flow conditions. Additional carbohydrate deformation bands near ~1184 $cm^{-1}$ further indicate enhanced structural polysaccharide organization within the root cell walls (**Fig. 5c**).[70] As seen already, the phenylalanine increases in FiP condition at ~1003 $cm^{-1}$ for the roots (**Fig. 5 b**). Phenylalanine is a key precursor of lignin biosynthesis through the phenylpropanoid pathway. Since lignin is a major structural component of plant cell walls, increased phenylalanine availability can contribute to enhanced root mechanical strength and structural stability.[71] The increase in carotenoid, phenylalanine, and carbohydrate bands indicates enhanced antioxidant capacity, metabolic activity, and cell-wall organization under flow. These biochemical changes are consistent with the structural adaptations observed in roots under flow conditions.

Notably, these biochemical adjustments extend beyond the root zone. As witnessed in **Fig. 5e-h**, shoot spectra show minimal differences between SiPP and SiP, indicating that confinement alone does not induce systemic metabolic changes. In contrast, FiP shoots exhibit stronger signals at ~957 $cm^{-1}$ and ~1003 $cm^{-1}$ (**Fig. 5e**). A pronounced increase is also observed at ~1523 $cm^{-1}$, corresponding to chlorophyll derivatives (**Fig. 5f**). Enhanced signals at ~1155 $cm^{-1}$ and ~1185 $cm^{-1}$ are associated with chlorophyll derivatives and tryptophan-related metabolites (**Fig. 5g**). [72,73] Elevated chlorophyll signatures suggest improved photosynthetic competence and photoprotective balance. Increased tryptophan signals point toward auxin-related metabolic modulation, reflecting integrated root-shoot signalling. Shoots also display a higher-wavenumber band near ~2673 $cm^{-1}$, corresponding to C-H overtone and combination vibrations of carbohydrates (**Fig. 5h**), indicating increased structural biomolecules.[70] Increased chlorophyll-related signals suggest improved photoprotective balance, while tryptophan features indicate modulation of auxin-related metabolism. Important to mention, only flow conditions produce these systemic effects underscores that advective hydrodynamics, rather than mere confinement, is the threshold stimulus for whole-plant physiological recalibration. On the whole, Raman signatures from roots and shoots reveal a consistent biochemical hierarchy (FiP > SiP > SiPP), reflecting the distinct physical environments experienced by the seedlings. Overall, confinement within the POD induces moderate biochemical adjustments without causing detrimental stress relative to petri plate growth, instead promoting soil-like priming responses such as structural and osmoprotective adaptations. The introduction of controlled flow further enhances pigment biosynthesis, structural stability, and metabolic coordination, establishing flow as the most favourable condition within the graded hydrodynamic environment of the POD (**Fig. 5**).

In concert with the morphological and anatomical results (**Figs. 3-4**), the Raman data establish a clear link between hydrodynamics and plant biochemistry, confirming the effectiveness of the POD as a soil-analogous platform with capability of eliciting physiologically meaningful whole-plant responses rather than stress-dominant effects.

**Systemic Physiological Consequences: Photosynthesis and Nitrogen Uptake**

The physiological consequences of hydrodynamic flow within the POD system were evaluated by examining both chlorophyll fluorescence responses and nitrogen uptake.[74,75] Chlorophyll fluorescence measurements revealed small but consistent differences between stagnant and flow conditions within the confined POD channels. The effective Photosystem II (PSII) quantum yield, Y(II), increased slightly from 0.807 (SiP) to 0.820 (FiP), indicating marginally improved photochemical utilization of absorbed light under flow. Correspondingly, the regulated non-photochemical dissipation, Y(NPQ), decreased from 0.143 (SiP) to 0.132 (FiP), suggesting reduced reliance on protective heat dissipation. The non-regulated energy dissipation, Y(NO), remained low and nearly unchanged (0.0497 - SiP; 0.048 - FiP), indicating

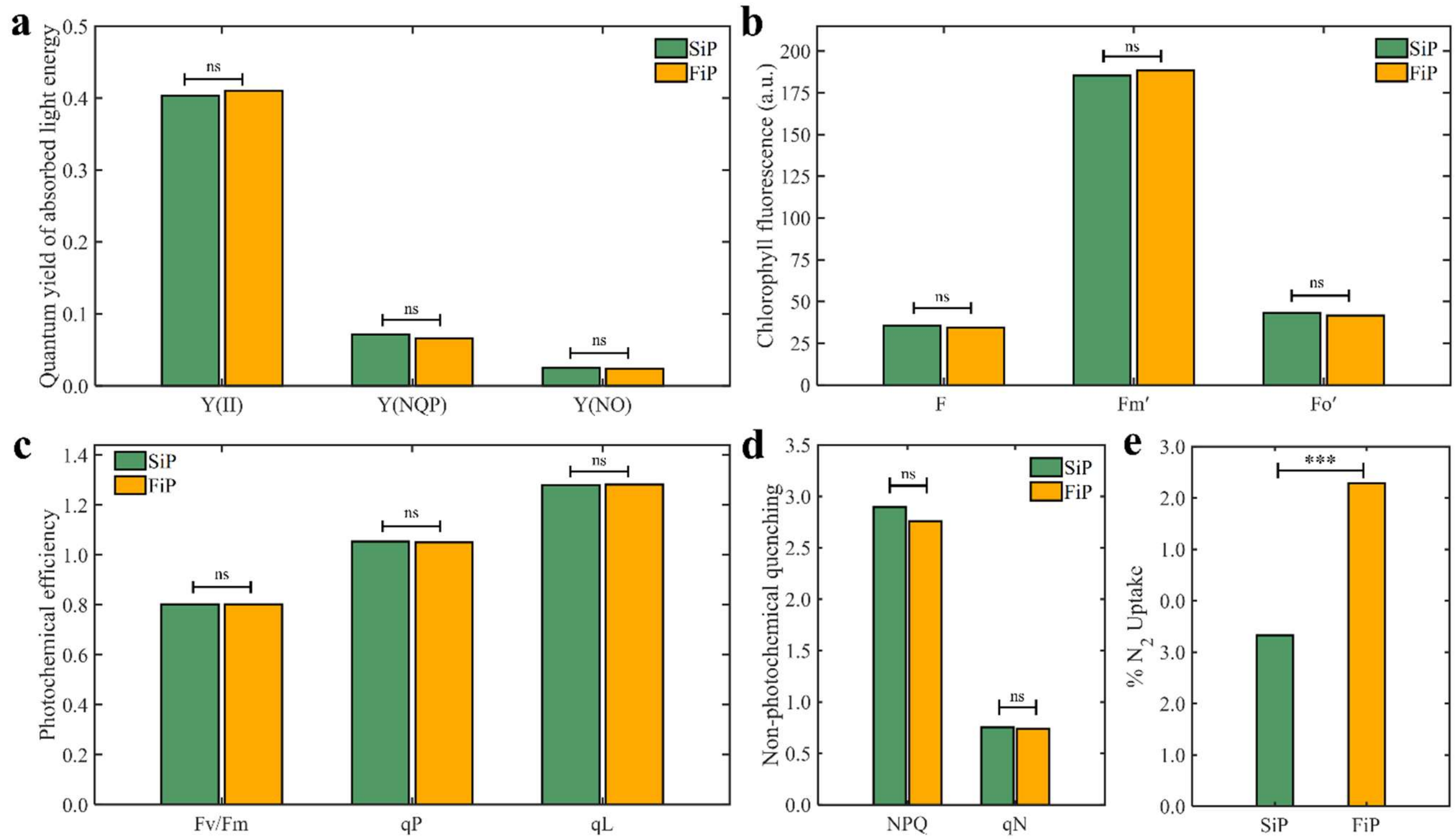

**Figure 6**: Chlorophyll fluorescence responses of seedlings under stagnant (SiP) and flow (FiP) conditions using PAM Fluorimeter. (A) Energy partitioning within photosystem II (PSII) showing the quantum yields of photochemistry Y(II), regulated non-photochemical energy dissipation Y(NPQ), and non-regulated energy loss Y(NO). (B) Light-adapted chlorophyll fluorescence parameters including steady-state fluorescence F, maximum fluorescence Fm′, and minimum fluorescence Fo′. (C) Photochemical efficiency parameters consisting of the maximum quantum efficiency of PSII Fv/Fm and photochemical quenching coefficients qP and qL. (D) Non-photochemical quenching parameters NPQ and qN, representing photoprotective heat dissipation mechanisms. Data (A-E) represent mean ± SD (n = 3). Differences between treatments were not statistically significant ($p > 0.05$). (E) Nitrogen uptake comparison between stagnant and flow conditions. Overall, hydrodynamic flow is associated with slightly higher PSII photochemical performance while reducing non-photochemical energy dissipation and significantly increasing nitrogen uptake compared with stagnant conditions. Data represent mean ± SD (n = 3). Statistical significance between flow and stagnant treatments was determined using Student's t-test (***$p < 0.001$).

-the absence of photodamage in either condition (**Fig. 6a**). Light-adapted fluorescence parameters showed minor shifts, with a small decrease in steady-state fluorescence (F) from 35.67 (stagnant) to 34.33 (flow) and minimum fluorescence in the light-adapted state (Fo′) from 43.33 to 41.67 under flow conditions, while maximum fluorescence in the light-adapted state (Fm′) increased slightly from 185.33 to 188.33 (**Fig. 6b**). These trends suggest a slightly higher proportion of open PSII reaction centres and improved energy utilisation. The maximum quantum efficiency (Fv/Fm) remained constant at 0.802 under both stagnant and flow conditions (**Fig. 6c**), indicating stable PSII performance. Photochemical quenching coefficients, qP and qL, were also similar between stagnant and flow conditions (**Fig. 6c**), indicating that the proportion of open PSII reaction centres and electron transport capacity were largely unaffected by flow.[74] Non-photochemical quenching (NPQ) decreased slightly from 2.896 under stagnant conditions to 2.760 under flow, while non-photochemical quenching coefficient (qN) decreased from 0.753 to 0.741, indicating reduced energy dissipation as heat and slightly better use of absorbed light (**Fig. 6d**).[74] Although the differences were not statistically significant (ns, $p > 0.05$), seedlings exposed to underlying flow hydrodynamics exhibited a slight increase in effective quantum yield Y(II) and a modest reduction in non-photochemical quenching Y(NPQ) with NPQ comparable to stagnant conditions.[75] This suggests that improved nutrient transport in roots under flow may subtly enhance photosynthetic efficiency without inducing measurable changes in PSII maximum efficiency (Fv/Fm). Quite notably, this observation shows consistency with the enhanced root elongation and structural remodelling observed under hydrodynamic flow (**Figs. 3-4**).[74] Albeit, nitrogen uptake showed a strong response to hydrodynamic transport. The mean nitrogen uptake increased from 1.66 (% by mass) under stagnant conditions to 3.14 (% by mass) under flow, representing nearly a two-fold increase (**Fig. 6e**). Under stagnant conditions (SiP), nutrient transport to confined roots occurs primarily by diffusion, which leads to the formation of a depletion of nutrient boundary layer around the root surface. In FiP, hydrodynamic flow reduces this boundary layer and continuously replenishes nutrients at the root interface, maintaining a higher concentration gradient and enabling greater nutrient flux toward the root. The presence of flow may also improve oxygen availability and support root metabolic activity involved in active nutrient transport.

Broadly, the results demonstrate that hydrodynamic flow within the POD device enhances nutrient acquisition while maintaining stable photosynthetic performance (**Fig. 6**). Although PSII photochemistry remains statistically unchanged between stagnant and flow conditions, flow slightly improves photochemical energy utilization and substantially increases nitrogen uptake. Together with the structural and biochemical responses discussed earlier, these observations highlight how the POD platform enables controlled investigation of plant responses under distinct physical regimes - FiP (confinement + flow + hydrostatic/normal stress), SiP (confinement + hydrostatic/normal stress), and SiPP (hydrostatic/normal stress only)-and demonstrates the physiological advantage of hydrodynamic nutrient transport within confined root environments. All together with earlier morphological (**Fig. 3**), anatomical (**Fig. 4**) and biochemical (**Fig. 5**) results, these findings show that hydrodynamic transport in the POD influences plant function across multiple scales.

**Mechanical Stresses: Numerical Modelling and Simulations**

Having established the graded biochemical responses across FiP, SiP, and SiPP conditions, we next undertake an effort to delve into the underlying physical factors contribute to internal mechanical stresses under loading. Numerical simulations were therefore performed to characterize pressure distribution and local internal mechanical stress distribution within the channels of POD (**Fig. 7**). These simulations help interpret the experimentally observed elongation bias, cortical remodelling, carotenoid enrichment, and enhanced nitrogen uptake (**Figs. 3-6**).

To distinguish effect of SiP and FiP conditions, hydrostatic pressure ($\rho gh$) was independently estimated to separate baseline compression from flow-driven stresses. The resulting normal or static pressure is 5.51 Pa under FiP and 4.89 Pa under SiP conditions, confirming that hydrostatic loading remains nearly identical between both conditions and is primarily determined by geometry and fluid height. Thus, flow primarily contributes to additional mechanical stresses.

The local internal mechanical stress decreases from CP1 to CP5, from $10^4$ to $10^2$ N/m$^2$, in the FiP condition (**Fig. 7 a**), which includes both hydrodynamic and static loading. In contrast, the decrease in local internal mechanical stress for the SiP condition is only one order of magnitude, from $10^1$ to $10^0$ N/m$^2$ (**Fig. 7 b**). As discussed earlier, the normal (static) pressure is nearly the same in both FiP and SiP conditions and mainly acts on the root surface. However, in the FiP condition, hydrodynamic loading generates an additional resistive shear stress that acts tangentially along the root periphery. This tangential loading is significant in FiP but absent in SiP condition. As a result, the maximum local internal mechanical stress in FiP is nearly three orders of magnitude higher than in SiP condition. It should also be noted that the local internal mechanical stress decreases from CP1 to CP5, i.e., from the plant inlet towards the root tip. The root region near the inlet is partly outside the nutrient-filled area and therefore does not experience hydrostatic loading. CP1 represents the location where the root first comes in contact with water and acts as a transition point between unloaded and loaded regions. As the static load increases beyond CP1, the root tends to minimize the resulting internal mechanical stress. This behaviour is similar to a cantilever beam under a uniformly distributed

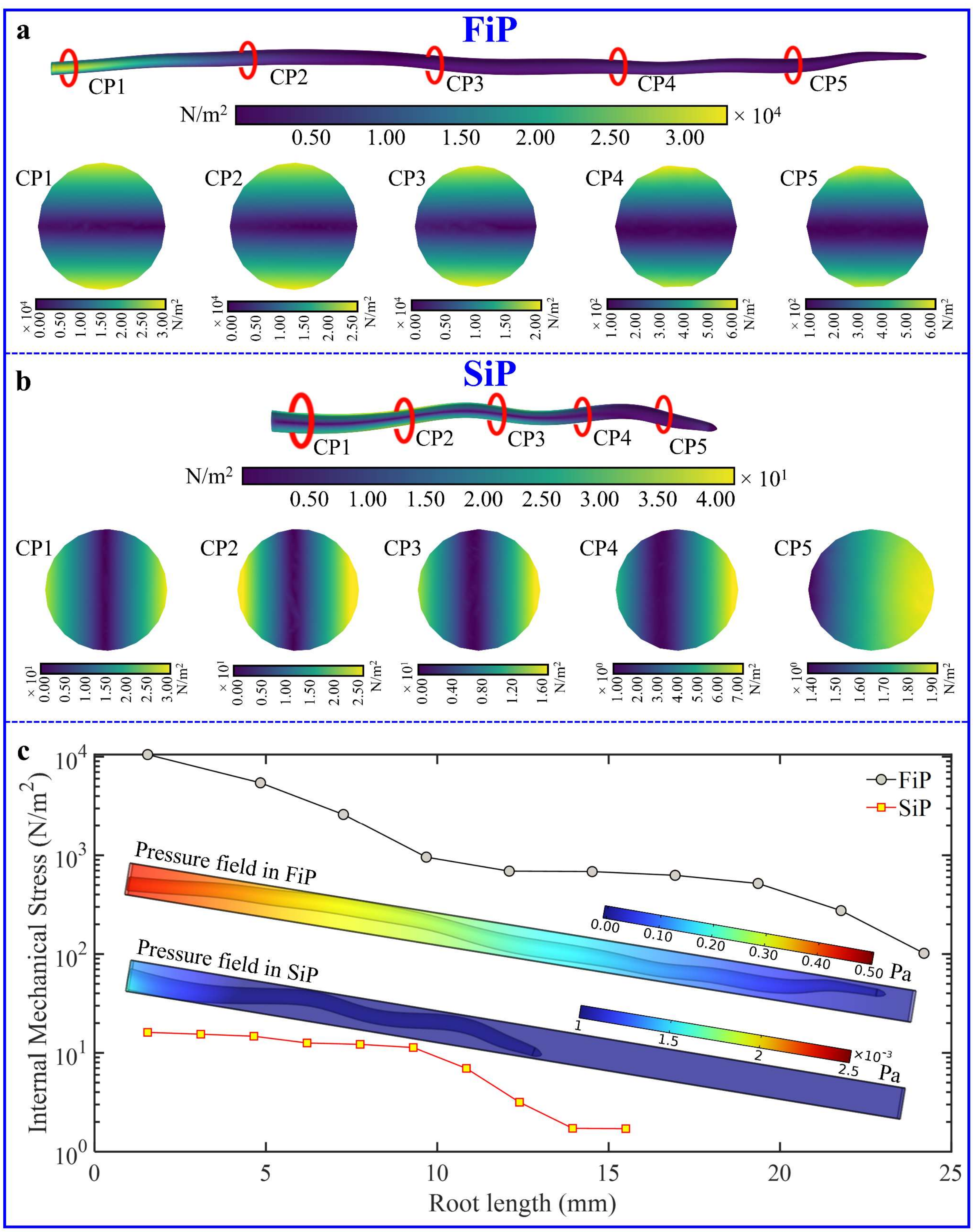

**Figure 7**: Spatial distribution of mechanical stress along the root under flow and stagnant conditions after 36 hrs. (a) Numerical simulations showing surface stress distribution along the root-channel interface under hydrodynamic flow (top row) and stagnant conditions (bottom row). Colour maps depict axial and cross-sectional stress contours at representative locations along the root. Under flow conditions, higher stresses are concentrated near the inlet and gradually decrease along the root length, whereas stagnant conditions exhibit substantially lower stresses throughout the root surface. (b) Quantitative variation of integrated surface-normal stress along root length. Flow (black) shows a steep decrease from inlet to outlet while remaining orders of magnitude higher than stagnant conditions (red) across all axial positions. The inset (log scale) highlights the two- to three-order-of-magnitude

difference in mechanical loading between flow and stagnant conditions. Overall, the figure shows that hydrodynamic flow increases mechanical loading within the POD platform.

load, where CP1 corresponds to the fixed end region that experiences the highest internal mechanical stress. The **Fig. 7 c** represents the section averaged internal mechanical stress which also follows the tread same as the local internal mechanical stress for both FiP and SiP conditions. The inset images in **Fig. 7c** show the pressure distribution along the channel in the FiP and SiP conditions. In FiP, the pressure decreases from the inlet region towards the root-tip, whereas the pressure remains nearly constant throughout the channel in SiP condition. It is to be noted that, the pressure magnitude is much higher in FiP condition, especially near the root inlet region. The FiP condition experiences both static and hydrodynamic loading, which produces a stronger pressure field along the root. In contrast, SiP is mainly governed by static loading and therefore shows a much weaker pressure field. The gradual reduction in pressure from the base towards the tip is consistent with the decrease in internal mechanical stress observed in **Fig. 7 a** and **b**.

While SiP conditions provide confinement under low-stress conditions, FiP conditions introduce strong shear-driven forces that act as mechanobiological stimuli. This separation highlights the role of the POD architecture in imposing controlled confinement while enabling tuneable flow-driven stresses. These stresses provide the physical basis for the morphological remodelling, biochemical enrichment, and enhanced physiological performance observed in roots grown within the POD platform (**Figs. 3-6**).

Quantitative morphometric indices further support this interpretation. Roots under FiP condition showed pronounced longitudinal elongation (**Figs. 3-4**). The cortical cell aspect ratio nearly doubled under flow compared to stagnant conditions, indicating a shift toward elongation-dominated growth. These results indicate that the stresses generated under flow are associated with elongation-dominated growth without evidence of adverse effects on root development. Mechanistically, SiP condition relies on diffusion-limited nutrient supply, restricting elongation while maintaining baseline cortical organization. In contrast, FiP condition introduces hydrodynamic shear together with enhanced nutrient uptake, which may contribute to the observed elongation-dominated growth (**Fig. 7**). Numerical simulations further show that the POD uniquely combines confinement with controllable hydrodynamic loading. Under FiP condition, roots experience a coupled confinement-flow-normal stress environment, whereas SiP condition retains confinement but lack shear-driven forcing. The increased phenylalanine level (**Fig. 5 b**) observed under FiP conditions for the roots also support structural adaptation, as phenylalanine serves as a precursor for lignin biosynthesis through the phenylpropanoid pathway, which contributes to cell wall strengthening and root structural stability. Similarly, the higher ROS level observed under FiP conditions (**Fig. 3c and d**) suggests increased physiological activity during root growth. Together with enhanced elongation and increased phenylalanine accumulation, these results indicate active cellular remodelling under hydrodynamic flow. The aforementioned discussion demonstrates that the POD provides physiologically relevant mechanobiological stimulation under moderate hydrostatic pressures.

## Conclusion

The Plant-on-a-Disc (POD) device, an eight-channel phytofluidic platform, is introduced in this study as a scalable and optically accessible system capable of replicating key physical features of the rhizosphere while enabling non-invasive, *in situ* analysis of living seedlings. By integrating geometric confinement with controlled hydrodynamic flow, the POD creates a microenvironment that reproduces coupled mechanical and transport processes analogous to those encountered by roots in natural soil systems. Dimensionless hydrodynamic analysis confirms that the device operates within a regime of creeping laminar flow and convection-dominated nutrient transport under physiologically safe conditions, thereby ensuring both biological relevance and analytical stability. Experimental results show that hydrodynamic flow strongly influences root development. Under flow conditions, roots exhibit faster axial elongation, higher ROS accumulation and anisotropic cellular expansion, characterized by elongated cortical cells and reduced radial thickening. In contrast, stagnant conditions promote radial proliferation, lesser accumulation of ROS and the formation of thicker roots, consistent with diffusion-limited nutrient transport and the absence of directional mechanical simulation. These structural changes are accompanied by biochemical responses, including increased cellulose-related and carotenoid signals, indicating enhanced structural organisation and elevated antioxidant capacity under flow-mediated conditions. Physiological analyses further reveal that photosynthetic performance remains stable under both conditions, with a slight improvement in energy utilisation under flow. Importantly, nitrogen uptake nearly doubles under flow regime, highlighting the critical role of advective transport in nutrient uptake. The numerical simulations support these observations, showing that fluid flow introduces significantly higher shear and normal stresses while maintaining nearly unchanged hydrostatic pressure distributions. These stresses act as directional mechanical cues that promote elongation-dominated growth patterns. In tandem, these results establish a direct mechanistic relationship between hydrodynamics and plant response across structural, biochemical, and physiological levels as established from both qualitative estimation, such as ROS generation, and quantitative analysis through numerical simulations. Within the POD environment, geometric confinement provides a controlled baseline environment, while hydrodynamic flow actively drives coordinated root remodelling and improved nutrient acquisition, thereby bridging the gap between simplified *in vitro* cultivation systems and realistic soil-like microenvironments. Beyond the biological insights obtained in this work, the POD also serves as a robust and versatile platform for plant science research. Its eight-channel architecture allows parallel experiments under uniform hydraulic conditions with full optical accessibility for multimodal imaging and spectroscopic analysis. This capability makes the system particularly suitable for studying root mechanobiology, nutrient transport dynamics, and plant-environment interactions, and stress adaptation under controlled environments. Comprehensively, the POD integrates principle of fluid mechanics and plant biology into a unified phytofluidic platform for studying soil-root interactions, with potential applications in high-throughput plant screening, nutrient optimisation strategies, and the development of climate-resilient agricultural technologies.

**Author Contributions**

K.A., Conceptualization, data curation, visualization, formal analysis, methodology, investigation, experimentation, writing - original draft; S.K.M., Formal analysis, visualization, data curation, methodology, writing - review & editing; P.K.M., Conceptualization, methodology, supervision, writing - review & editing, project administration, funding acquisition, resources.

**Conflicts of interest**

There are no conflicts to declare.

**Data availability**

All data supporting the findings of this study are available within the article and its Supplementary Information file. Additional raw data are available from the corresponding author upon reasonable request.

**Acknowledgements**

This work is supported by the Science and Engineering Research Board (SERB), Govt. of India, under the project no. CRG/2022/000762. The Ph.D. fellowship of Mr. Kaushal Agarwal is supported by the Ministry of Education (MoE), Govt. of India. The authors gratefully acknowledge Mr. Rameshwar Shukla, PhD scholar, Department of Biosciences and Bioengineering, Indian Institute of Technology Guwahati, Guwahati 781039, India, for his valuable assistance in acquiring the initial experimental readings. The authors thank the TIDF (TIH) Indian Institute of Technology, Guwahati for their active support. The authors are grateful to CIF, IIT Guwahati for the instrumentation facility.

**References**

1 E. Kolb, M. Quiros, G. J. Meijer, M. B. Bogeat-Triboulot, A. Carminati, E. Andò, L. Sibille and F. Anselmucci, in *Soft Matter in Plants*, The Royal Society of Chemistry, 2022, pp. 165–202.

2 H. Zhang, X. Yan, M. Zhang, Y. Zhao, S. Jiang, Y. Jiang, Y. Wei, Y. Zhang and L. Sun, *J. Agric. Food Chem.*, 2025, **73**, 30895–30903.

3 A. Pierret and C. J. Moran, in *Encyclopedia of Earth Sciences Series*, eds. J. Gliński, J. Horabik and J. Lipiec, Springer Netherlands, Dordrecht, 2011, vol. Part 4, pp. 628–632.

4 E. Zelazny and G. Vert, *Plant Physiol.*, 2014, **166**, 500–508.

5 K. Robe and M. Barberon, *Curr. Opin. Plant Biol.*, 2023, **74**, 102376.

6 J. B. Morgan and E. L. Connolly, *Nat. Educ. Knowl.*, 2013, **4**, 2.

7 J. Alonso-Serra, *New Phytol.*, 2026, **249**, 722–728.

8 A. A. Cardoso, M. T. Andrade, E. R. Bucior and S. C. V. Martins, *Plant Physiol.*, 2025, **199**, kiaf521.

9 J. Joshi, B. D. Stocker, F. Hofhansl, S. Zhou, U. Dieckmann and I. C. Prentice, *Nat. Plants*, 2022, **8**, 1304–1316.

10 A. Sampathkumar, A. Yan, P. Krupinski and E. M. Meyerowitz, *Curr. Biol.*, 2014, **24**, R475-83.

11 M. Delarue, *npj Biol. Phys. Mech.*, 2025, **2**, 8.

12 P. J. Gregory, *Physiol. Determ. Crop yield*, 1994, 65–93.

13 B. K. Pandey, T. S. George, H. V. Cooper, C. J. Sturrock, T. Bennett and M. J. Bennett, *J. Exp. Bot.*, 2025, **76**, 1500–1509.

14 J. P. Lynch, C. F. Strock, H. M. Schneider, J. S. Sidhu, I. Ajmera, T. Galindo-Castañeda, S. P. Klein and M. T. Hanlon, *Plant Soil*, 2021, **466**, 21–63.

15 R. Shen, B. Borer, D. Ciccarese, M. M. Salek and A. R. Babbin, *mSphere*, 2024, **9**, e0018524.

16 W. Cao, G. Yan, H. Hofmann and A. Scheuermann, *Geotechnics*, 2025, **5**, 2.

17 A. G. Hunt and M. Sahimi, *Rev. Geophys.*, 2017, **55**, 993–1078.

18 P. Voothuluru, Y. Wu and R. E. Sharp, *Plant Cell*, 2024, **36**, 1377–1409.

19 R. Karlova, D. Boer, S. Hayes and C. Testerink, *Plant Physiol.*, 2021, **187**, 1057–1070.

20 A. Erktan, D. Or and S. Scheu, *Soil Biol. Biochem.*, 2020, **148**, 107876.

21 P. Mehra, J. Banda, L. L. P. Ogorek, R. Fusi, G. Castrillo, T. Colombi, B. K. Pandey, C. J. Sturrock, D. M. Wells and M. J. Bennett, *Annu. Rev. Plant Biol.*, 2025, **76**, 467–492.

22 H. Zhang, X. Yan, M. Zhang, Y. Zhao, S. Jiang, Y. Jiang, Y. Wei, Y. Zhang and L. Sun, *J. Agric. Food Chem.*, 2025, **73**, 30895–30903.

23 P. R. Ryan, E. Delhaize, M. Watt and A. E. Richardson, *Ann. Bot.*, 2016, **118**, 555–559.

24 J. Moran and C. McGrath, *Biotechniques*, 2021, **71**, 605–614.

25 A. Muhammad, Z. U. Khan, J. Khan, A. S. Mashori, A. Ali, N. Jabeen, Z. Han and F. Li, *Front. Plant Sci.*, 2025, **16**, 1638675.

26 X. Zhao, L. Zhai, J. Chen, Y. Zhou, J. Gao, W. Xu, X. Li, K. Liu, T. Zhong, Y. Xiao and X. Yu, *J. Agric. Food Chem.*, 2024, **72**, 15401–15415.
27 L. D. Cohen, E. Moratto and C. E. Stanley, *Lab Chip*, 2026, **26**, 1273–1298.
28 C. F. Kaiser, A. Perilli, G. Grossmann and Y. Meroz, *J. Exp. Bot.*, 2023, **74**, 3851–3863.
29 N. Frey, U. M. Sönmez, J. Minden and P. LeDuc, *Nat. Commun.*, 2022, **13**, 3195.
30 S. Panja, S. K. Mehta, J. Kalita, M. K. Prasad and P. K. Mondal, *Phys. Fluids*, 2024, **36**, 111914.
31 K. Agarwal, S. K. Mehta and P. K. Mondal, *Lab Chip*, 2024, **24**, 3775–3789.
32 P. Padhi, S. K. Mehta, K. Agarwal and P. K. Mondal, *Phys. Fluids*, 2024, **36**, 043602.
33 S. K. Mehta, A. Talukdar, S. Panja, J. Kalita, S. Wongwises and P. K. Mondal, *Soft Matter*, 2025, **21**, 1269–1285.
34 S. Panja, S. K. Mehta, J. Kalita, D. Panchal, X. Zhang and P. K. Mondal, *J. Colloid Interface Sci.*, 2026, **715**, 140281.
35 C. E. Stanley, J. Shrivastava, R. Brugman, E. Heinzelmann, D. van Swaay and G. Grossmann, *New Phytol.*, 2018, **217**, 1357–1369.
36 G. Grossmann, W. J. Guo, D. W. Ehrhardt, W. B. Frommer, R. V. Sit, S. R. Quake and M. Meier, *Plant Cell*, 2011, **23**, 4234–4240.
37 P. M. Mafla-Endara, C. Arellano-Caicedo, K. Aleklett, M. Pucetaite, P. Ohlsson and E. C. Hammer, *Commun. Biol.*, 2021, **4**, 889.
38 M. Zhu, C. W. Hsu, L. L. Peralta Ogorek, I. W. Taylor, S. La Cavera, D. M. Oliveira, L. Verma, P. Mehra, M. Mijar, A. Sadanandom, F. Perez-Cota, W. Boerjan, T. M. Nolan, M. J. Bennett, P. N. Benfey and B. K. Pandey, *Nature*, 2025, **642**, 721–729.
39 L. Ma, Y. Shi, O. Siemianowski, B. Yuan, T. K. Egner, S. V. Mirnezami, K. R. Lind, B. Ganapathysubramanian, V. Venditti and L. Cademartiri, *Proc. Natl. Acad. Sci. U. S. A.*, 2019, **166**, 11063–11068.
40 H. Priks, T. Butelmann, A. Illarionov, T. G. Johnston, C. Fellin, T. Tamm, A. Nelson, R. Kumar and P. J. Lahtvee, *ACS Appl. Bio Mater.*, 2020, **3**, 4273–4281.
41 O. Ali, I. Cheddadi, B. Landrein and Y. Long, *New Phytol.*, 2023, **238**, 62–69.
42 L. Hoermayer, J. C. Montesinos, N. Trozzi, L. Spona, S. Yoshida, P. Marhava, S. Caballero-Mancebo, E. Benková, C. P. Heisenberg, Y. Dagdas, M. Majda and J. Friml, *Dev. Cell*, 2024, **59**, 1333-1344.e4.
43 V. Hernández-Hernández, D. Rueda, L. Caballero, E. R. Alvarez-Buylla and M. Benítez, *Front. Plant Sci.*, 2014, **5**, 265.
44 D. C. Trinh, J. Alonso-Serra, M. Asaoka, L. Colin, M. Cortes, A. Malivert, S. Takatani, F. Zhao, J. Traas, C. Trehin and O. Hamant, *Curr. Biol.*, 2021, **31**, R143–R159.
45 Y. Yan, Z. Sun, P. Yan, T. Wang and Y. Zhang, *New Phytol.*, 2023, **239**, 1609–1621.
46 A. Sampathkumar, *Dev.*, 2020, **147**, dev177964.
47 V. Mirabet, P. Das, A. Boudaoud and O. Hamant, *Annu. Rev. Plant Biol.*, 2011, **62**, 365–385.
48 M. A. Andersen and J. Schouenborg, *Sci. Rep.*, 2023, **13**, 1–12.
49 I. Miranda, A. Souza, P. Sousa, J. Ribeiro, E. M. S. Castanheira, R. Lima and G. Minas, *J. Funct. Biomater.*, 2022, **13**, 2.
50 S. H. Song, C. K. Lee, T. J. Kim, I. C. Shin, S. C. Jun and H. Il Jung, *Microfluid. Nanofluidics*, 2010, **9**, 533–540.
51 D. Mandal, S. Datta, G. Raveendar, P. K. Mondal and R. Nag Chaudhuri, *Plant J.*, 2023, **113**, 106–126.
52 M. Kafi and Z. Rahimi, *Soil Sci. Plant Nutr.*, 2011, **57**, 341–347.
53 L. G. Angelini, C. Clemente and S. Tavarini, *Agric.*, 2021, **11**, 937.
54 R. J. McMurtrey, *Tissue Eng. Part C. Methods*, 2016, **22**, 221–249.
55 C. M. Vera and M. V. Mendoza, *Brazilian J. Chem. Eng.*, 2025, **42**, 1143–1158.

56 M. M. Naderi, H. Gao, J. Zhou, I. Papautsky and Z. Peng, *Lab Chip*, 2025, **25**, 2874–2886.
57 J. X. J. Zhang and K. Hoshino, in *Micro and Nano Technologies*, eds. J. X. J. Zhang and K. B. T.-M. S. and N. (Second E. Hoshino, Academic Press, 2019, pp. 113–179.
58 B. Khatoon, Shabih-Ul-Hasan and M. S. Alam, in *14 International Symposium on Process Systems Engineering*, eds. Y. Yamashita and M. B. T.-C. A. C. E. Kano, Elsevier, 2022, vol. 49, pp. 691–696.
59 H. G. Matthies and J. Steindorf, *Comput. Struct.*, 2003, **81**, 805–812.
60 S. Pabi, M. K. Khan, S. K. Jain, A. K. Sen and A. Raj, *Phys. Fluids*, 2023, **35**, 101908.
61 M. Rahmani, A. Hammouti and A. Wachs, *Phys. Fluids*, 2018, **30**, 043301.
62 C. Truesdell and W. Noll, eds. C. Truesdell, W. Noll and S. S. Antman, Springer Berlin Heidelberg, Berlin, Heidelberg, 2004, pp. 1–579.
63 K. C. Maturi, I. Haq and A. S. Kalamdhad, *Environ. Sci. Pollut. Res.*, 2022, **29**, 84600–84615.
64 N. Kennard, R. Stirling, A. Prashar and E. Lopez-Capel, *Agronomy*, 2020, **10**, 1092.
65 D. L. Gelardi, I. H. Ainuddin, D. A. Rippner, J. E. Patiño, M. A. Najm and S. J. Parikh, *Soil*, 2021, **7**, 811–825.
66 J. Jehlička and A. Oren, *Front. Microbiol.*, 2013, **4**, 380.
67 T.-F. Hsieh, K.-J. Yu and S.-Y. Lin, *Dis. Markers*, 2007, **23**, 705630.
68 A. Jarin, U. K. Ghosh, M. S. Hossain, A. Mahmud and M. A. R. Khan, *Discov. Agric.*, 2024, **2**, 127.
69 W. Z. Payne and D. Kurouski, *Front. Plant Sci.*, 2021, **11**, 616672.
70 E. Wiercigroch, E. Szafraniec, K. Czamara, M. Z. Pacia, K. Majzner, K. Kochan, A. Kaczor, M. Baranska and K. Malek, *Spectrochim. Acta - Part A Mol. Biomol. Spectrosc.*, 2017, **185**, 317–335.
71 J. El-Azaz, B. Moore, Y. Takeda-Kimura, R. Yokoyama, M. Wijesingha Ahchige, X. Chen, M. Schneider and H. A. Maeda, *Nat. Commun.*, 2023, **14**, 7242.
72 C. Zhou, J. R. Diers and D. F. Bocian, *J. Phys. Chem. B*, 1997, **101**, 9635–9644.
73 Z.-L. Cai, H. Zeng, M. Chen and A. W. D. Larkum, *Biochim. Biophys. Acta - Bioenerg.*, 2002, **1556**, 89–91.
74 E. H. Murchie and T. Lawson, *J. Exp. Bot.*, 2013, **64**, 3983–3998.
75 M. A. Naseer, S. Hussain, A. Mukhtar, Q. Rui, G. Ru, H. Ahmad, Z. Q. Zhang, L. B. Shi, M. S. Asad, X. Chen, X. B. Zhou and X. Ren, *Front. Plant Sci.*, 2024, **15**, 1396929.